\documentclass[prl,aps,amssymb,twocolumn,superscriptaddress,showkeys,notitlepage,nobalancelastpage]{revtex4-1}

\usepackage{graphicx}
\usepackage{dcolumn}
\usepackage{bm}
\usepackage{ulem}
\usepackage{braket} 
\usepackage[caption=false]{subfig} 
\usepackage{changepage}
\usepackage{float}
\usepackage{amsmath}
\usepackage{xcolor}
\usepackage{bbold}

\usepackage{dsfont} 

\usepackage{hyperref}
\hypersetup{colorlinks=true}

\begin{document}

\title{Axion Topology in Photonic Crystal Domain Walls}

\author{Chiara Devescovi*}
\email{chiara.devescovi@dipc.org}
\affiliation{Donostia International Physics Center, Paseo Manuel de Lardizabal 4, 20018 Donostia-San Sebastian, Spain.}

\author{Antonio Morales-P\'erez*}
\affiliation{Donostia International Physics Center, Paseo Manuel de Lardizabal 4, 20018 Donostia-San Sebastian, Spain.}

\author{Yoonseok Hwang}
\affiliation{Department of Physics, University of Illinois at Urbana-Champaign, Urbana, IL, USA}

\author{Mikel García-Díez}
\affiliation{Donostia International Physics Center, Paseo Manuel de Lardizabal 4, 20018 Donostia-San Sebastian, Spain.}
\affiliation{Physics Department, University of the Basque Country (UPV/EHU), Bilbao, Spain}

\author{Iñigo Robredo}
\affiliation{Max Planck Institute for Chemical Physics of Solids, Dresden D-01187, Germany}
\affiliation{Donostia International Physics Center, Paseo Manuel de Lardizabal 4, 20018 Donostia-San Sebastian, Spain.}

\author{Juan Luis Mañes}
\affiliation{Physics Department, University of the Basque Country (UPV/EHU), Bilbao, Spain}

\author{Barry Bradlyn}
\affiliation{Department of Physics, University of Illinois at Urbana-Champaign, Urbana, IL, 
USA}

\author{Aitzol Garc\'ia-Etxarri}
\email{aitzolgarcia@dipc.org }
\affiliation{Donostia International Physics Center, Paseo Manuel de Lardizabal 4, 20018 Donostia-San Sebastian, Spain.}
\affiliation{IKERBASQUE, Basque Foundation for Science, Mar\'ia D\'iaz de Haro 3, 48013 Bilbao, Spain.}

\author{Maia G. Vergniory}
\email{maia.vergniory@cpfs.mpg.de}
\affiliation{Max Planck Institute for Chemical Physics of Solids, Dresden D-01187, Germany}
\affiliation{Donostia International Physics Center, Paseo Manuel de Lardizabal 4, 20018 Donostia-San Sebastian, Spain.}

\date{\today}

\begin{abstract}

Axion insulators are 3D magnetic higher-order topological insulators protected by inversion-symmetry that exhibit hinge-localized chiral channels and induce quantized topological magnetoelectric effects. 
Recent research has suggested that axion insulators may be capable of detecting dark-matter axion-like particles by coupling to their axionic excitations. Beyond its fundamental theoretical interest, designing a photonic AXI offers the potential to enable the development of magnetically-tunable photonic switch devices through the manipulation of the axionic modes and their chiral propagation using external magnetic fields. 
Motivated by these facts, in this work, we propose a novel approach to induce axionic band topology in gyrotropic 3D Weyl photonic crystals gapped by supercell modulation. To quantize an axionic angle, we create domain-walls across inversion-symmetric photonic crystals, incorporating a phase-obstruction in the supercell modulation of their dielectric elements. This allows us to bind chiral channels on inversion-related hinges, ultimately leading to the realization of an axionic chiral channel of light. Moreover, by controlling the material gyrotropic response, we demonstrate a physically accessible way of manipulating the axionic modes through a small external magnetic bias, which provides an effective topological switch between different 1D chiral photonic fiber configurations. Remarkably, the unidirectional axionic hinge states supported by the photonic axion insulator are buried in a fully connected 3D dielectric structure, thereby being protected from radiation through the electromagnetic continuum. As a result, they are highly suitable for applications in guided-light communication, where the preservation and non-reciprocal propagation of photonic signals are of paramount importance.

\end{abstract}

\maketitle

\section{Introduction}
Axion insulators (AXIs) \cite{wilczek1987two,fu2007topological,hughes2011inversion,turner2012quantized,fang2012bulk,wieder2018axion,qi2009inducing,sekine2021axion,wu2016quantized,varnava2018surfaces,sekine2021axion,gonzalez2022axion,zhang2013surface,yue2018symmetry,varnava2018surfaces,khalaf2018higher,xu2019higher,olsen2020gapless,tanak2020theory,takahashi2020bulk,gong2022half} are 3D inversion ($\mathcal{I}$)-symmetric magnetic higher-order topological insulators (HOTIs)\cite{schindler2018higher1,schindler2018higher2} 
which induce various topological magnetoelectric effects, such as the quantized magneto-optical Faraday and Kerr rotation, the image magnetic monopole effect, and half-quantized surface Hall conductance \cite{qi2009inducing,sekine2021axion,wu2016quantized,varnava2018surfaces}. 
The topological properties of AXIs arise from the quantization of their electromagnetic coupling term, the so-called topological $\theta$-angle \cite{wilczek1987two,fu2007topological,hughes2011inversion}, which is pinned to $\pi$ in presence of $\mathcal{I}$-symmetry (or other $\theta$-odd operations such as roto-inversions and  time-reversed rotations).
 \cite{varnava2020axion}. 

AXIs are of significant interest because, acting as HOTI, they are able to support hinge-localized chiral modes, which propagate in the form of unidirectional axionic channels \cite{sehayek2020charge,you2016response,gooth2019axionic}. 
These hinge-states are expected to emerge at the 1D facets of an AXI crystallite or in the presence of 1D dislocations in the AXI lattice, where gradients of the $\theta$ angle arise, inducing the formation of axionic strings \cite{varnava2018surfaces,wang2013chiral,you2016response}. 
Recent studies \cite{varnava2018surfaces, olsen2020gapless} have shown that the chiral propgation of the AXI hinge-localized modes is highly-tunable. 
Especially in the presence of a ferromagnetic order, it is possible to switch between different hinge-modes configurations via external magnetic control, allowing magnetic re-routing of conducting channels from one input into one or more outputs. In the context of Photonic Crystals (PhCs), this remarkable property of AXIs could allow for devices that manipulate, direct and deviate the 1D non-reciprocal flow of light, with relevant applications for optical communication technologies and for the development of magnetically-tunable photonic switch devices. Until now, no proposals have been presented for axion-based PhCs or axion-protected light propagation.
Furthermore, recent studies have suggested the use of AXI materials for detecting axion-like particles, that constitute dark-matter candidates \cite{chigusa2021axion,marsh2019proposal,li2010dynamical}. This is due to the fact that emergent axionic excitations in AXI couple with electromagnetism, $\mathcal{L} \propto  \theta \mathbf{E}\cdot\mathbf{B}$, similar to the axion-photon coupling observed in high-energy physics for light dark-matter, which follows $\mathcal{L} \propto a \mathbf{E}\cdot\mathbf{B}$.  In PhCs, photons can interact with external magnetic fields via gyrotropy, they display a non-zero effective mass,  
and they are wavelength-tunable via lattice size-scaling, all of which are essential ingredients for the realization of an axion haloscope \cite{chigusa2021axion,millar2017dielectric,lawson2019tunable,yokoi2018stimulated}. The demonstration of an AXI in a PhC could represent an opportunity to bridge these two different approaches in the study of axion-photon coupling.

Despite the theoretical significance and potential applications of AXIs, no proposals have yet been put forward for their implementation in PhCs: our work aims to propose and demonstrate the first theoretical model and general design strategy for photonic AXIs in 3D PhCs. To induce a photonic axionic band topology, we incorporate a phase-obstruction in the Supercell Modulation (SM) of the dielectric elements within gyrotropic Weyl PhCs\cite{devescovi2021cubic,devescovi2022vectorial}. The SM is designed as an $\mathcal{I}$-symmetric, static, geometric deformation of the PhC lattice, enabling a experimental implementation of the PhC without necessitating any dynamic driving. Serving as a photonic analog of a charge-density-wave (CDW) \cite{wieder2020axionic,gooth2019axionic,shi2021charge,fukuyama1978dynamics}, the SM couples Weyl points with opposite topological charges while maintaining the $\mathcal{I}$-symmetry of the model. 

The resulting AXI is dubbed \textit{relative}, because it is only exhibited at the interface of two PhCs 
with a quantized \textit{relative} axionic angle $\delta\theta$ and vanishing \textit{relative} Chern numbers. This approach is grounded in the concept that a dislocation of the CDW phase in a specific class of $\mathcal{I}$-symmetric Weyl semimetals (WS) acts as a dynamic axion field \cite{wang2013chiral,wieder2020axionic,gooth2019axionic}. Consequently, the domain-wall separating the phase-obstructed CDW-WS can be interpreted as the critical point between a trivial insulator and an AXI. By employing this strategy, we successfully realize a photonic \textit{relative} axion insulator (\textit{r}AXI) in a realistic gyrotropic setup.

By inserting planar dislocations in the dielectric modulation phase, we bind 1D chiral channels on $\mathcal{I}$-related hinges, that provide a PhC realization of an axion domain wall protected by $\mathcal{I}$-symmetry.

Remarkably, the 1D channels supported by the  PhC are buried in a fully connected 3D dielectric structure, thus protected from radiation through the electromagnetic continuum \cite{joannopoulos2008molding}. This design not only represents the first instance of a tunable HOTI with chiral hinge states in 3D PhCs \cite{kim2020recent}, but the observed 1D-modes are also consistent with a single, unidirectional axionic channel that wraps around the central phase-obstructed core, endowing the photonic hinge-channels with non-reciprocal propagation properties. 

Lastly, we propose a physically viable method for manipulating these axionic hinge modes by controlling the PhC gyrotropic response using a small external magnetic bias. Specifically, we induce gyrotropy-induced transitions in the photonic AXI, which function as an efficient topological switch between various 1D photonic fiber configurations. Interestingly, recent experimental advancements in 3D gyrotropic crystals have demonstrated that imparting a magnetic response to 3D photons is possible, with a high degree of control and intensity \cite{liu2022topological,xi2023topological}. These findings suggest that it may be possible to manipulate, direct, and deviate the 1D non-reciprocal flow of light in a photonic AXI using state-of-the-art experimental setups.  The capability of manipulating the HOTI hinge states in the photonic AXI via gyrotropy underscores the potential of the proposed design for creating magnetically-tunable photonic switch devices, thereby paving the way for advancements in axion-based photonics.

The main body of the manuscript is divided into three sections: In Section \textcolor{red}{1}, we provide the bulk design and topological characterization of the photonic \textit{relative} axion insulator (\textit{r}AXI); In Section \textcolor{red}{2}, we show how to make \textit{r}AXI topology manifest by creating a domain-wall between phase-obstructed $\mathcal{I}$-symmetric \textit{r}AXIs; In Section \textcolor{red}{3}, we demonstrate how to generate and manipulate the higher order topology of the PhC, by controlling the chiral propagation of the axionic channels of light. Finally, in the Methods Section, we show how to efficiently simulate the electromagnetic response of the \textit{r}AXI, via a transversality-enforced tight-binding model (TETB) \cite{antonio2023}. This model is capable of capturing and regularizing the $\Gamma$-point electromagnetic obstruction that arises in 3D PhCs, due to the transversality constraint of the Maxwell equations \cite{antonio2023,christensen2022location}.

\normalsize

\section{Results}

\subsection{1. Relative axion topology}
\label{sec:bulk}
\begin{figure}[h]
\centering
\includegraphics[width=75mm]{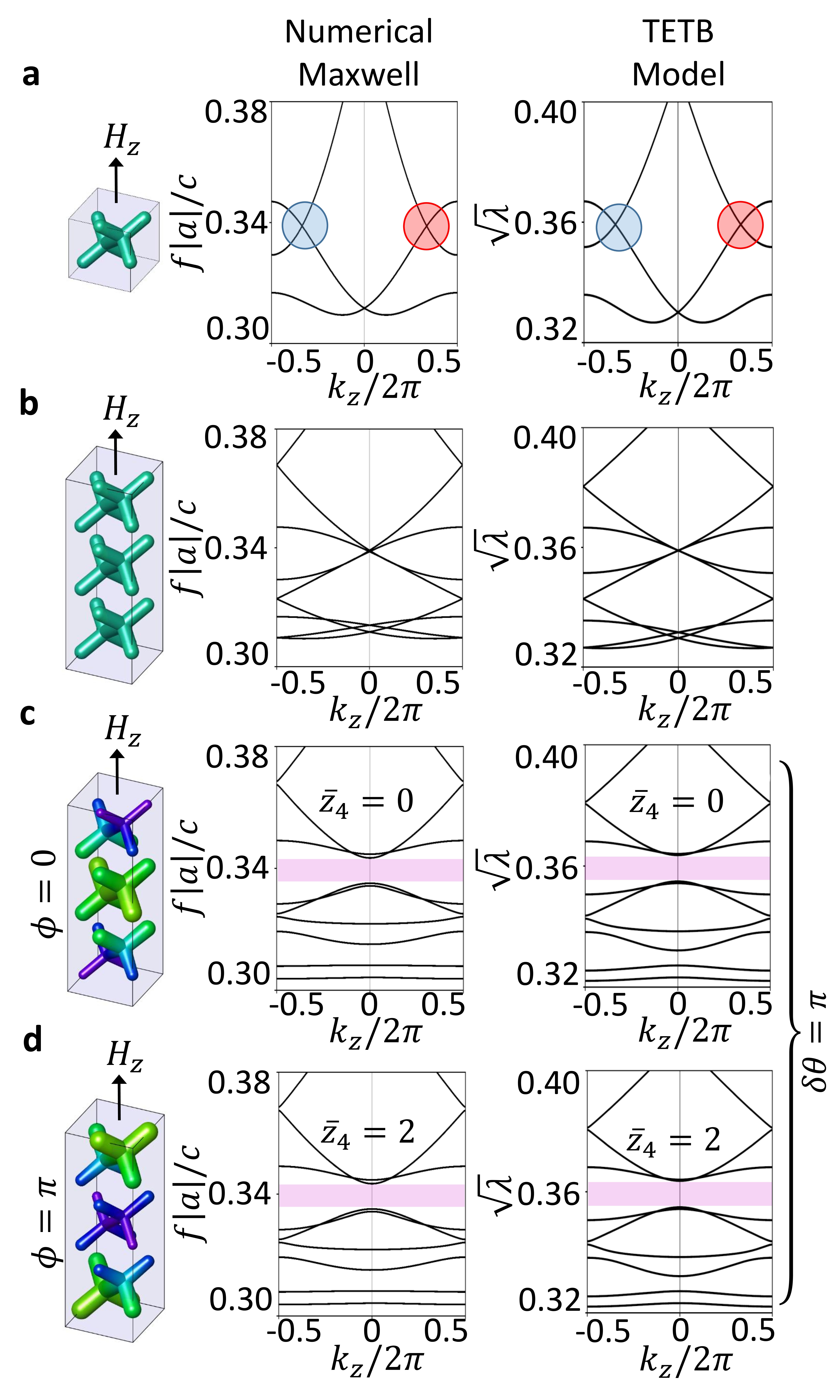}
\captionsetup[figure]{font=small}
\caption{Effect of a supercell modulation on the photonic Weyl bands. Left panels show the PhC geometry and the reduced frequencies $f|a|/c$, where $c$ is the speed of light and $|a|$ the scale invariant lattice parameter, obtained by solving numerically the Maxwell equations. Right panels show the square-root of the transversality-enforced tight-binding model (TETB) eigenvalues $\sqrt{\lambda}$, consistent with the mapping between the Schrödinger and electromagnetic wave equations, that relates energies and frequencies quadratically ($\lambda \sim\omega^2$, see \cite{antonio2023}). In these plots we show the $k_z$ line, for fixed $k_x=\pi$ and $k_y=\pi$. Weyl points which are separated by a $|\mathbf{Q}|=2\pi/N$ distance in momentum space as in panel (a), are superimposed on an artificial supercell in panel (b), and then coupled by a commensurate SM of period  $N=2\pi/|\mathbf{Q}|$ as in panels (c-d). The supercell amplitude is $r_m/r_0=1/20$ for the PhC and $V_{4c}=-V_{4b}=1/150$ for the TETB. Panel (c) and (d) differ solely for the angular phase of the supercell modulation $\phi$, with $\phi=0$ in panel (c), and $\phi=\pi$ in panel (d). }
\label{weyl}
\end{figure}

Our starting setup for inducing photonic AXI band topology consists of an $\mathcal{I}$-symmetric gyrotropic PhC \cite{yang2017weyl,devescovi2022vectorial,antonio2023} under an external magnetic field  $\textbf{H}=(0,0,H_z)$, as shown in Fig. \ref{weyl}\textcolor{red}{(a)}. 

In the presence of a gyroelectric medium the external magnetic field induces an off-diagonal imaginary component  in the permittivity tensor \cite{haldane2008possible,raghu2008analogs}, as expressed by the following equation:
\begin{equation}\label{eq:gyro}
{\varepsilon}_{\eta_z}=\left(\begin{array}{ccc}
\varepsilon_\perp & i\eta_{z} & 0 \\
-i \eta_{z} & \varepsilon_\perp & 0 \\
0 & 0 & \varepsilon
\end{array}\right),
\end{equation}
where $\eta_{z}=\eta_{z}(H_z)$ is the bias-dependent gyroelectric parameter with $\varepsilon_{\perp}=\sqrt{\varepsilon^2+\eta_z^2}$ and $\varepsilon$ the dielectric constant (here $\varepsilon=16$). As a consequence of time-reversal symmetry (TRS) breaking due to gyrotropy, a photonic Weyl dipole is generated in the Brillouin zone, along the direction of the $H_z$ magnetic field, as shown in Fig. \ref{weyl}\textcolor{red}{(a)}.
In the case of this dielectric lattice realization, the Weyl dipole separation increases proportionally to the external $H_z$ and can be magnetically tuned.

In order to emulate the effect of a CDW in condensed-matter systems \cite{gooth2019axionic,shi2021charge,sehayek2020charge} to open a topological gap, we introduce a $z$-directed SM of lattice period $N\in\mathds{N}$ and $N\geq2$, commensurate with the Weyl dipole separation $\mathbf{Q}=\mathbf{q}_{+}-\mathbf{q}_{-}$, where $\mathbf{q}_\pm$ are the locations of Weyl points with chirality $\pm 1$ in Brillouin zone.
Accordingly, we fix the Weyl points of opposite topological charge at approximately $\mathbf{q}_\pm=(\pi,\pi,\pi\pm \pi/N)$. This results in a folding of the BZ, as shown in Fig. \ref{weyl}\textcolor{red}{(b)}, and couples the Weyl points to open a non-trivial gap, as shown in Fig. \ref{weyl}\textcolor{red}{(c)}.

The SM is introduced as a local deformation $\delta r$ of the radius $r$ of the dielectric rods, according to the relation: $$\delta r=r_m \text{cos} (2\pi z/N|a| +\phi),$$ where $|a|$ is the lattice parameter of the starting photonic crystal, while $r_m$ and $\phi$ control, respectively, the amplitude and the phase of the dielectric modulation. Note that this represents a static geometric deformation of the PhC structure which can be stably implemented during the fabrication process and does not require any dynamical driving. The $\phi$ phase of the SM is the fundamental design parameter which we will set in order to induce axionic band topology.

\begin{figure}[h]
\centering
\includegraphics[width=50mm]{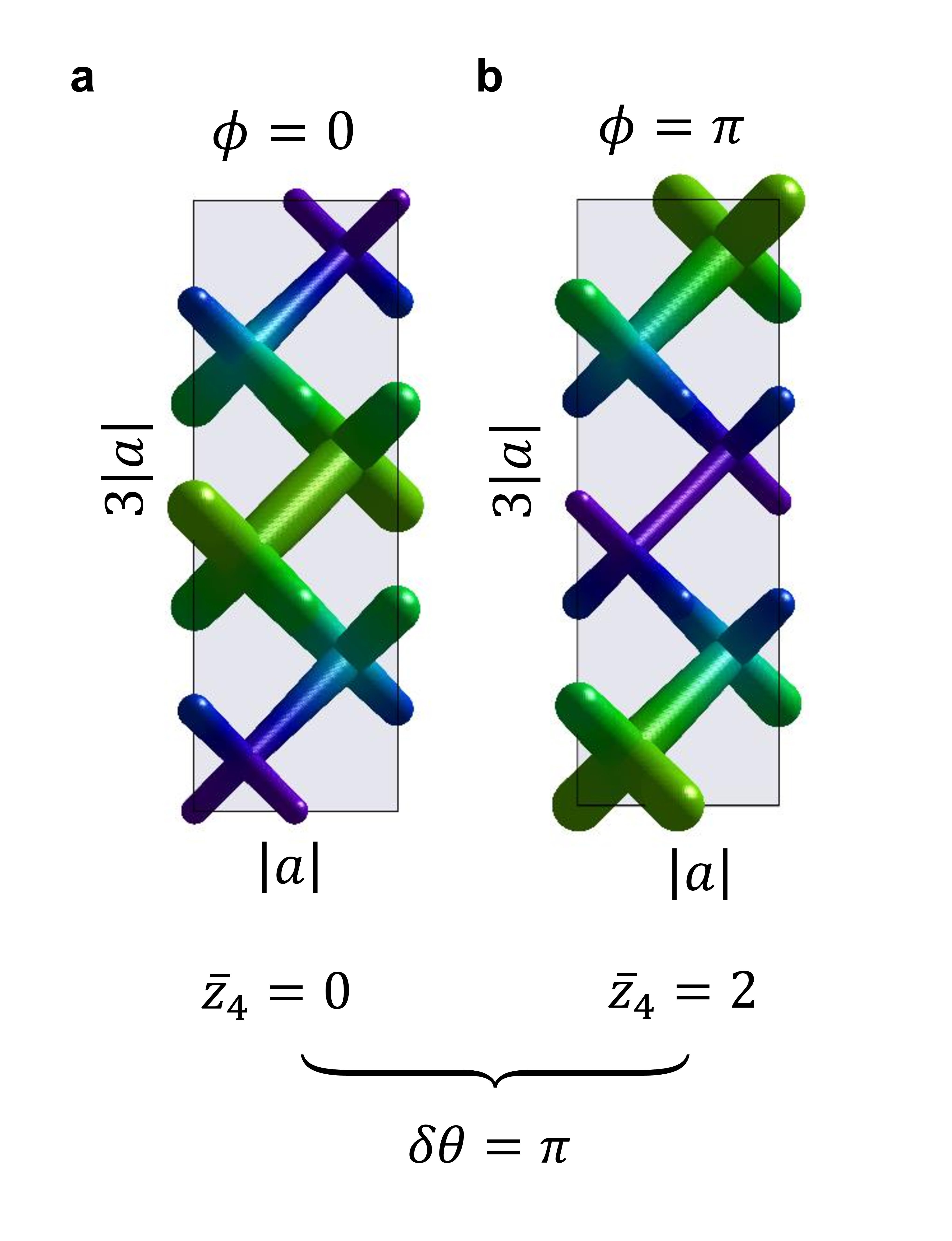}
\captionsetup[figure]{font=small}
\caption{3D  photonic \textit{r}AXI resulting from a $N=3$ periodic SM imposed on a gyrotropic Weyl photonic semimetal (side view). The SM acts as a local deformation of the diameter of the dielectric rods. The $z$-directed modulation is along the magnetization axis. The SM is centered at the inversion center of the unperturbed lattice: panels (a) and (b) correspond to an angular phase $\phi$ of the SM of $0$ and $\pi$, respectively. }
\label{geometry}
\end{figure}

For the purpose of preserving the $\mathcal{I}$-symmetry of the unperturbed PhC of Fig. \ref{weyl}\textcolor{red}{(a)}, which is crucial for axion behaviour, we pin the modulation at the $\mathcal{I}$-center and target only two specific values of the SM phase: $\phi=0$ and $\phi=\pi$. The corresponding modulated dielectric structures are shown in Fig. \ref{weyl}\textcolor{red}{(c-d)} in a 3D rendering, and in Fig. \ref{geometry} in a side view, for a $N=3$ modulation period. We see that both $\phi=0$ and $\phi=\pi$ phases display the same insulating spectrum. However, we will now demonstrate that their 3D photonic bulk gaps exhibit a different topological obstruction in the $\mathcal{I}$-symmetry-indicators associated to the quantization of their \textit{relative} axion angle $\delta\theta$.

To efficiently model the electromagnetic response of the PhC, we develop an analytical model of the 3D photonic bulk bands via the transversality-enforced tight-binding model method (TETB) introduced by Ref. \cite{antonio2023}. 
The TETB model is constructed via the introduction of  $\mathbf{v}^L$ auxiliary longitudinal modes, able to regularize the $\Gamma$-point obstruction arising from the transversality constraint of the Maxwell equations, as proposed in Ref. \cite{christensen2022location}. The positive-energy solutions of the TETB are mapped to frequency dispersion of the  $\mathbf{v}^T$  transverse electromagnetic modes, obtained by numerically solving the Maxwell equation via the MIT Photonic Bands package (MPB) \cite{johnson2001block}. The SM is introduced in the TETB via a simple onsite supercell-modulated potential, that mimics the local electromagnetic energy redistribution in the modulated dielectric rods (see Methods section): 

\begin{equation}\label{eq:mod_m}
    H_{\Delta}(\textbf{r},\textbf{H}) = H(\textbf{r},\textbf{H}) + \sum_i V_i\cos\left(\frac{2\pi z_i}{N|a|}+\phi\right)c_i^\dagger(\textbf{r}) c_i(\textbf{r}),
\end{equation}
where $H(\textbf{r},\textbf{H})$ is the real-space TETB Hamiltonian for the magnetic system before modulation, $|a|$ is the lattice parameter of the crystal before modulation, and $V_i$ and $\phi$ parameterize the amplitude and the phase of the modulation, respectively. Note that the sum in Eq. \eqref{eq:mod_m} runs over all the basis pseudo-orbitals used in the TETB model. As shown in the Methods section, the TETB reproduces all the bulk properties of the supercell-modulated PhC; a comparison of the respective bands and topology is displayed in Fig. \ref{weyl}.

In order to understand the role of $\mathcal{I}$-symmetry in protecting the \textit{r}AXI topology, we compute the magnetic symmetry-indicators (SI) $\nu^T_{\phi}=\{\bar{z}_{2,x},\bar{z}_{2,y},\bar{z}_{2,z}|\bar{z}_4\}$ \cite{watanabe2018structure,kim2019glide,takahashi2020bulk,xu2020high,elcoro2021magnetic,po2020symmetry} for the tranverse-electromagnetic modes of the PhC (the $\bar{\nonumber\nonumber}$ overbar stands for magnetic and the $(\nonumber)^{T}$ superscript indicates transverse bands). In particular, we focus our interest on the $\bar{z}_4$ strong index, which is associated to axion topology \cite{po2020symmetry,tanak2020theory,wieder2020axionic}.

As shown in the Methods section, we obtain, correspondingly for the two structures at $\phi=0$ and $\phi=\pi$:
\begin{equation}
    \nu^T_{\phi=0}=\{0,0,1|0\}
\end{equation}
and
\begin{equation}
    \nu^T_{\phi=\pi}=\{0,0,1|2\}
\end{equation}
where the $\delta \bar{z}_4=2$ discontinuity of the $\textit{even}$ $\bar{z}_4$ index stands to indicate a \textit{relative}  axionic obstruction. On the other hand, the invariance of the $\bar{z}_{2,z}$ term is related to an odd $C_{z}$ Chern invariant, which, as confirmed via photonic Wilson loop \cite{blanco2020tutorial,de2019engineering,devescovi2021cubic} calculations, is $C_{z}=1$ identically for both structures. Note that although we have computed $\nu^T_{\phi=0}$ and $\nu^T_{\phi=\pi}$ using the TETB model 
, the difference
\begin{equation} \label{eq:zdiff}\nu^T_{\phi=\pi}-\nu^T_{\phi=0} = \{0,0,0|2\}
\end{equation}
depends only on the sign of the modulation-induced band gap. We thus find that Eq.~\eqref{eq:zdiff} holds for the PhC.

To verify the quantization of the \textit{relative} axion angle between $\phi=0,\pi$, we compute the layer Chern number $G_z$ of a $z$-slab with its normal along the magnetization axis. As demonstrated in \cite{varnava2018surfaces,varnava2020axion}, a non-zero quantized axionic phase $\theta$ will manifests as an offset in $G_z$, 
according to the relation:
\begin{equation}
\label{layer_chern}
    G_z=C_z n_z+\theta/\pi
\end{equation}
where $n_z$ counts the layers of the slab.
Via this equation, we are able to extract the $\theta$ axion angle, from the $C_z$ Chern number of a single layer. 
\begin{figure}[h]
\centering
\includegraphics[width=70mm]{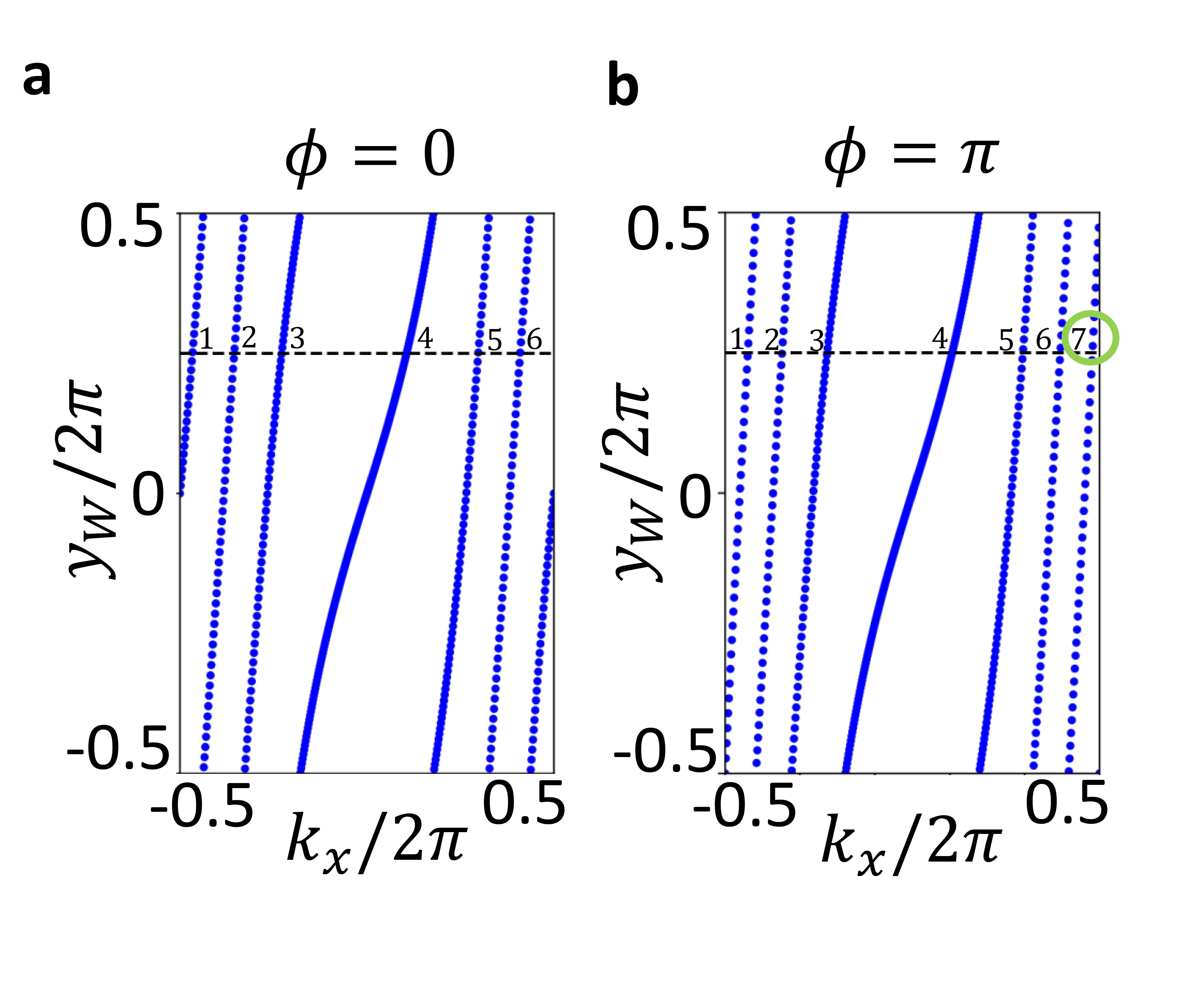}
\captionsetup[figure]{font=small}
\caption{Layer Wilson loop for a $z$-slab at $\phi=0$ (panel a) and $\phi=\pi$ (panel b) with $n_z=6$ layers. The $y$ Wannier energy centers wind respectively $n_z$ and $n_z+1$ times along $k_x$, with the $+1$ discontinuity shown in the green circle. This confirms a $\delta\theta=\delta\phi=\pi$ difference in the \textit{relative} axion angle. }
\label{layer}
\end{figure}

The slab Wilson loops, shown in Fig. \ref{layer}, wind $n_z$ and $n_z+1$ times, respectively for $\phi=0$ and $\phi=\pi$, confirming a $\delta\theta=\pi$ discontinuity in the axion $\theta$ angle. Therefore the $\phi=0,\pi$ supercell-modulated PhC represent \textit{r}AXI.

\subsection{2. Phase-obstructed domain walls}
\label{sec:surface}
\begin{figure}[h]
\centering
\includegraphics[width=80mm]{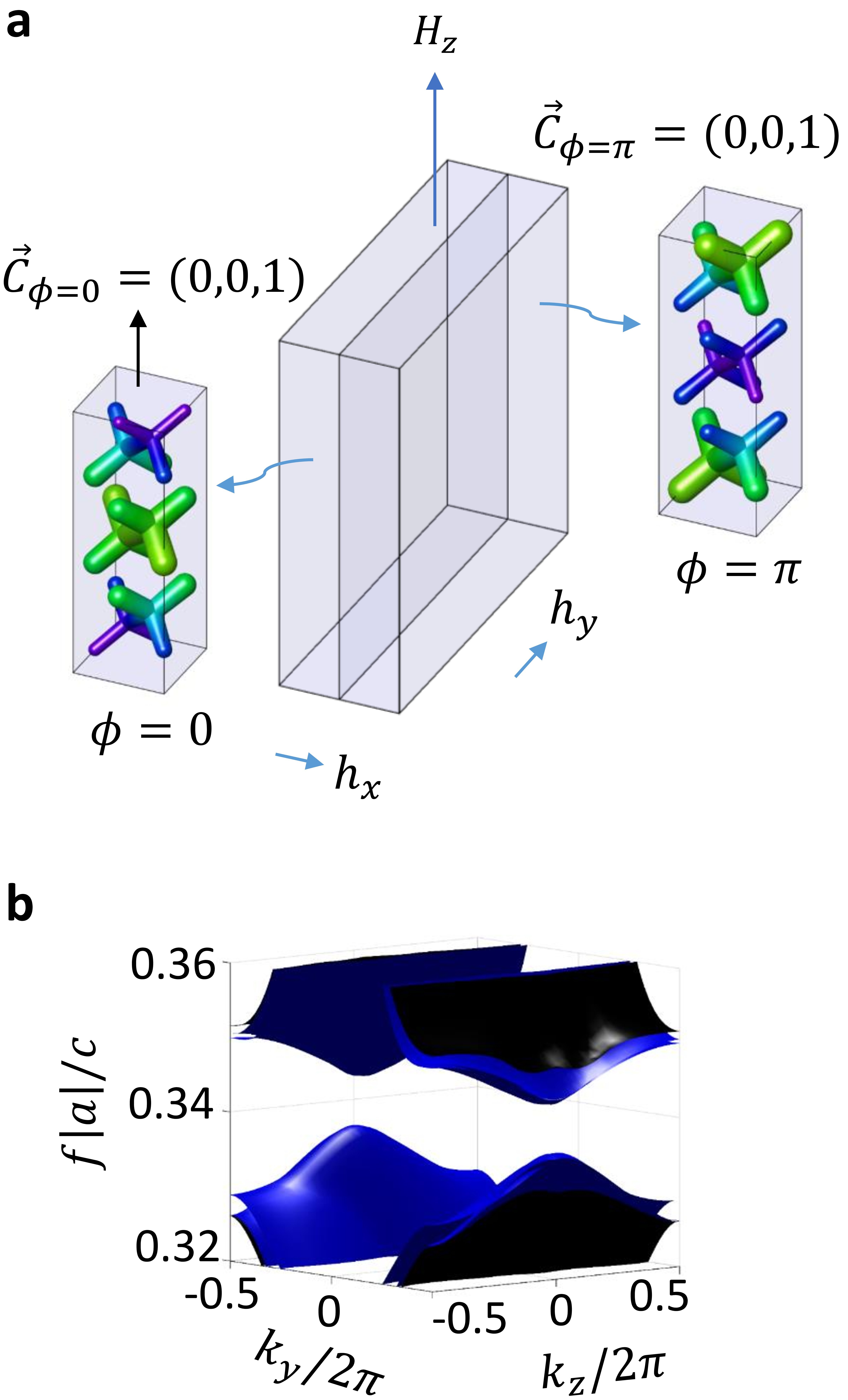}
\captionsetup[figure]{font=small}
\caption{Axionic surface gap for an $\mathcal{I}$-symmetric domain-wall with $\delta\theta\equiv\delta\phi=\pi$. In panel (a), PhC geometry of the phase-obstructed domain-wall configuration.  In panel (b), domain-wall band structure on the $x=0$ plane, with projected bulk bands in black, and surface-localized states in blue. }
\label{surface}
\end{figure}

In this section, our goal is to manifest the \textit{relative} axion topology. To accomplish this, we create a domain-wall in $x$ between the photonic 3D insulator with $\phi=0$ and its obstructed counterpart with $\phi=\pi$, i.e. imposing a \textit{relative} axion phase difference of $\delta\theta\equiv\delta\phi=\pi$, as shown in Fig. \ref{surface}\textcolor{red}{(a)}.  

We expect this domain-wall configuration to be formally equivalent to the critical point between an AXI with $\theta=\pi$ and a trivial insulator \cite{wang2013chiral,wieder2020axionic,wang2013chiral,gooth2019axionic,teo2010topological} and therefore gapped.
To ensure a surface gap, we apply a small tilt to the $z$-directed magnetic field, with $$\textbf{H}=(|h|\text{cos}(\sigma),|h|\text{sin}(\sigma),H_z)$$ and $|h|\ll|H_z|.$
As shown in the Supplementary Notes, the component of the magnetic perturbation normal to the interface plane ensures the existence of a surface gap, which is essential for the observation of the higher-order topology of the \textit{r}AXI. The tilted external field couples to the PhC, inducing an in-plane gyrotropic perturbation $\eta_{x,y}=\eta_{x,y}(h_{x,y})$ in the permittivity tensor. As result, as shown in Fig. \ref{surface}\textcolor{red}{(b)}, the PhC domain-wall bands are gapped. 

The size of the surface gap can be controlled via the $h_{x,y}$ bias, by gradually deviating from the gapless condition which results from the boundary condition choice, as demonstrated in the Supplementary Notes. 
In what follows we choose a boundary condition where the size of the surface gap vanishes in absence of any magnetization orthogonal to the interface plane: this boundary configuration is reached by maintaining the rod geometry continuously connected across interface for the PhC. In the TETB, this correspond to a surface potential that linearly interpolates between the two modulations.

Importantly, the $\phi=0$ and $\phi=\pi$ structures differ only in their $\bar{z}_4$ index but have an identical Chern vector. It is critical to maintain the condition of equal Chern vectors across the interface in order to prevent anomalous Hall surface states to populate the surface gap, consistently with vectorial bulk-boundary correspondence \cite{devescovi2022vectorial}.

\subsection{3. Gyrotropy-induced switching of HOTI states}
\label{sec:hinge}

\begin{figure}[h]
\centering
\includegraphics[width=70mm]{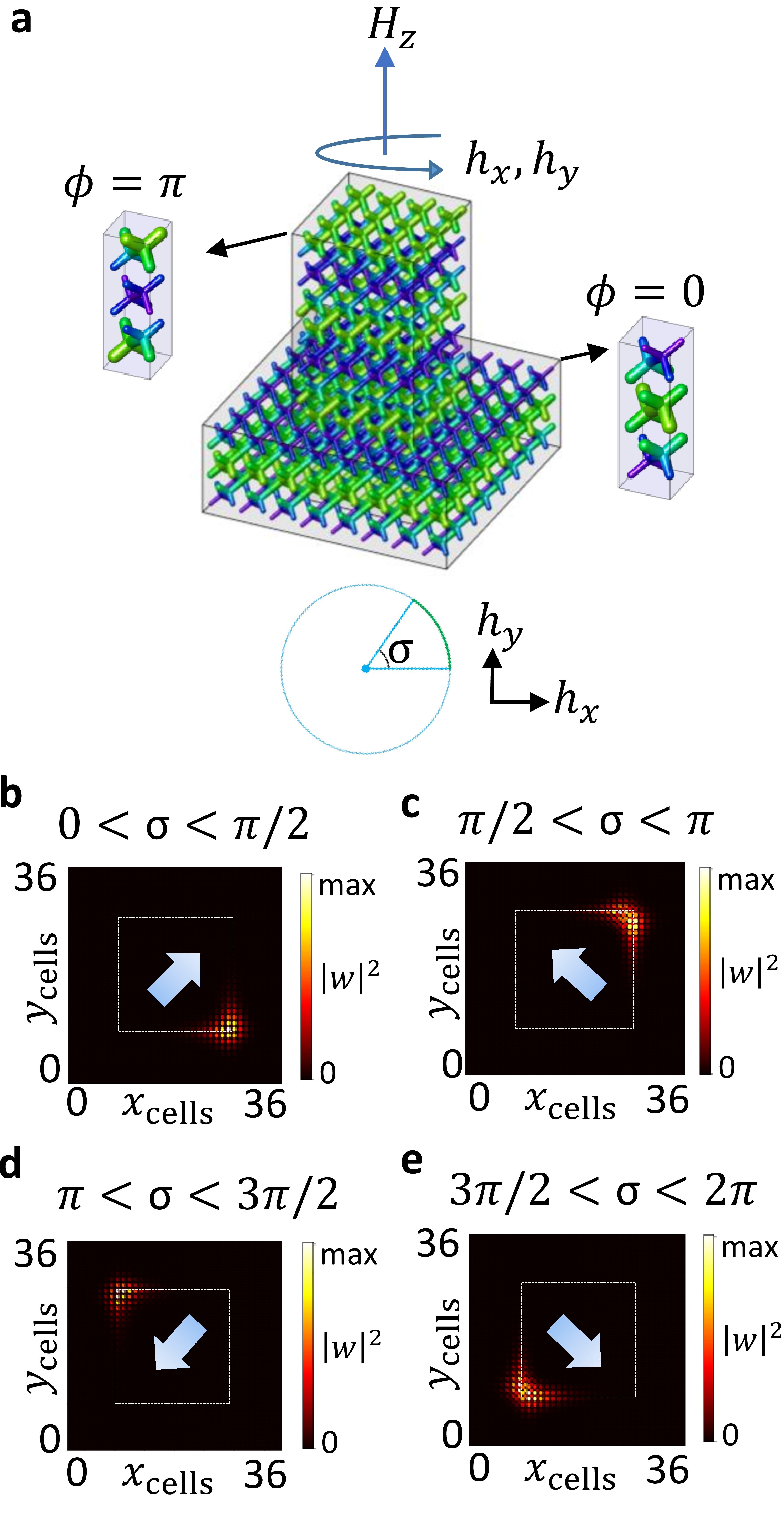}
\captionsetup[figure]{font=small}
\caption{Tunable AXI hinge states at $\Gamma$, for different magnetic bias configurations, computed via the TETB. Panel (a) displays the corresponding PhC dielectric structure. For visual purposes, the central $\phi=0$ core is extruded vertically with respect to the phase-obstructed embedding with $\phi=\pi$. Panels (b-e) correspond to $\alpha,\gamma,\beta,\delta$ configurations.  A single eigenvector is plotted here, upwards moving. The activation of the $90^\circ$-rotated hinges is made possible via a $h_{x,y}$ in-plane small bias component.}
\label{tunable}
\end{figure}

Next, to generate and manipulate a chiral hinge channel of light, we will be investigating the higher-order topology of the PhC. For this purpose, we construct an $\mathcal{I}$-symmetric $z$-wire configuration, embedding a $N_x$ $\times$ $N_y$ core of $\phi=0$ PhC, inside a $2N_x$ $\times$ 2$N_y$ region of PhC of the same material with $\phi=\pi$. The corresponding dielectric structure, which is fully connected, is shown in Fig. \ref{tunable}\textcolor{red}{(a)}, with the central rod extruded upwards, for better visualization. To keep the simulations affordable, we compute the boundary modes for this rod-geometry via the use of the TETB model. 

\begin{figure}[h]
\centering
\includegraphics[width=75mm]{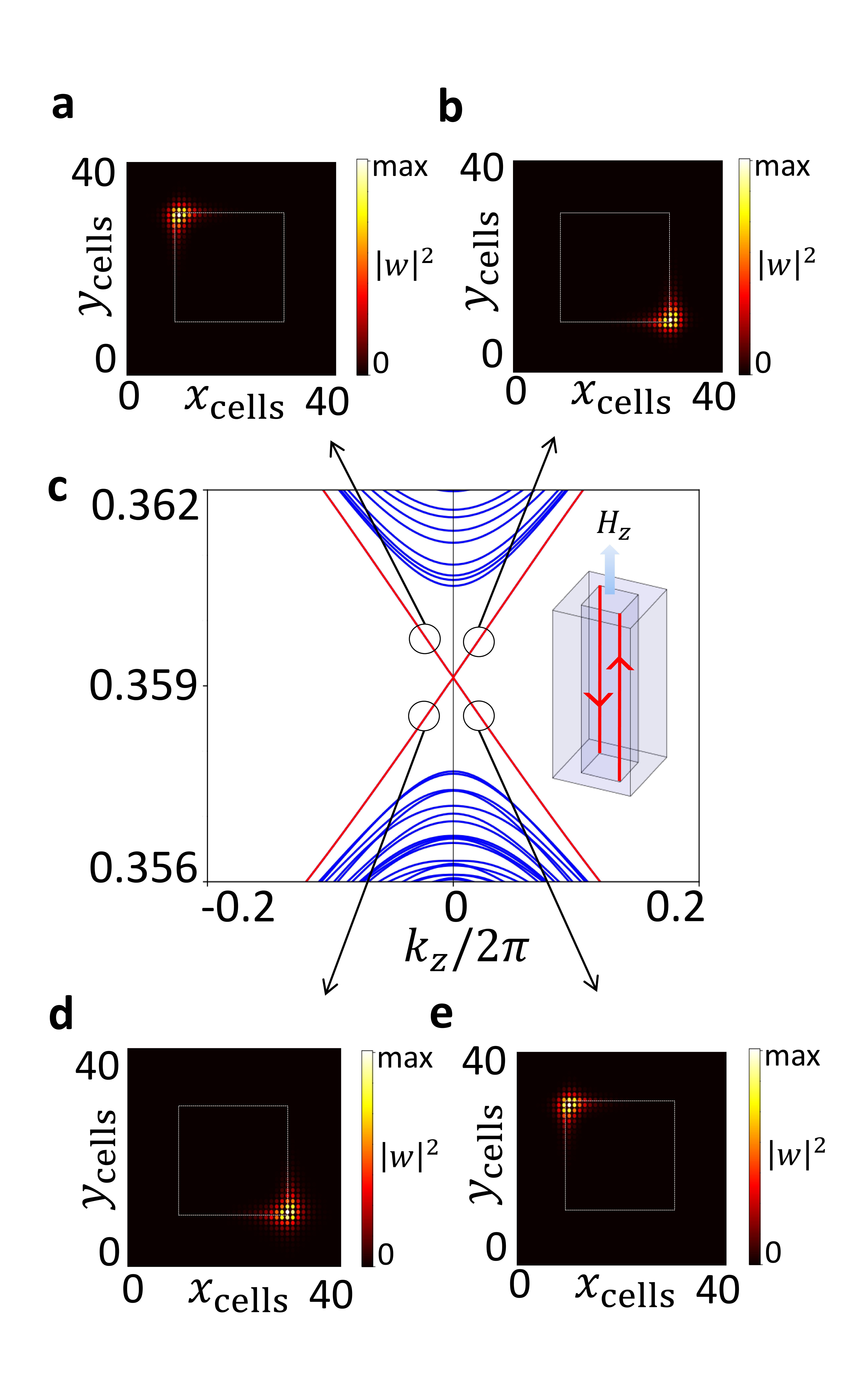}
\captionsetup[figure]{font=small}
\caption{Gapless AXI hinge states evaluated a $z$-wire configuration, with $2N_x$ $\times$ $2N_y=40 \times 40$ cells. The crystal structure is fully connected but presents an axion phase discontinuity of $\delta\theta\equiv\delta\phi=\pi$.  Projected surface bands in blue, hinge bands in red, in panel (c). The chiral modes are localized on $\mathcal{I}$-related hinges: a $xy$-cross section of the $z$-wire geometry is shown in panels (a,b,d,e). The flipping of the external $H_z$ field results in an overall exchange of the group velocity signs. These HOTI states are consistent with the existence of a single unidirectional mode wrapping around a central phase-obstructed core.}
\label{hinge}
\end{figure}

As shown in Fig. \ref{hinge}\textcolor{red}{(c)}, chiral gapless modes emerge as in-gap states in the projected domain-wall bands, consistent with the bulk–hinge correspondence of the photonic \textit{r}AXI. These HOTI states are consistent with the existence of a single unidirectional mode wrapping around a central phase-obstructed core. Moreover, their group velocity can be easily switched by flipping of the external magnetic bias $H_z$. Displayed for a cross-section of the connected structure in Fig. \ref{hinge}\textcolor{red}{(a-e)}, the 1D channels localize on $\mathcal{I}$-related hinges parallel to the $z$ direction.  

It is noteworthy that not all of the four $\mathcal{I}$-related hinges support chiral modes at once. Instead, the localization on either a pair of $\mathcal{I}$-related hinges or the other can be chosen by rotation of the small $h_{x,y}$ bias in the $xy$ plane, leading to 4 possible realizations of the hinges, $\alpha,\beta$ (with occupancy of the hinges passing through the corners on the $1\bar{1}0$ diagonal)  and $\gamma,\delta$ (with occupancy of the hinges passing through the corners on the $110$ diagonal), as shown in Fig. \ref{tunable}\textcolor{red}{(b-e)}. These different hinge-state configurations are plotted in Fig. \ref{tunable} at the $\Gamma$ point for the upwards-moving state. As shown in the Supplementary Notes, they can be regarded as distinct boundary-obstructed phases~\cite{khalaf2021boundary,wong2022higher}, since a surface gap (but not a bulk gap) must close in passing from one configuration to another.

The $\alpha$,$\beta$,$\gamma$,$\delta$ gyrotropic-bias-field induced transitions offer a promising and physically accessible way to manipulate the photonic 1D modes, via rotation of the PhC gyrotropic axis through magnetic control by external field. Therefore, the present platform can provide an effective photonic topological switch between different 1D photonic fiber configurations. 

Remarkably, the observed hinge modes are embedded within a fully connected 3D dielectric structure, making them highly suitable for guided-light communication applications, as they are protected from radiation through the electromagnetic continuum \cite{joannopoulos2008molding}.

By proposing the first tunable HOTI chiral hinge states in PhCs \cite{kim2020recent}, we provide a PhC realization and a distinct manifestation of the axionic hinge states predicted in supercell-modulated Weyl semimetals 
\cite{you2016response,wieder2020axionic,wang2013chiral,gooth2019axionic,teo2010topological}.

More specifically, the hinge modes of Fig. \ref{hinge}\textcolor{red}{(b-e)} are consistent with the presence of a single, unidirectional axionic mode wrapping around a central phase-obstructed core \cite{khalaf2018symmetry}.

\section{Conclusions}
In conclusion, we proposed a novel design strategy to induce axionic band topology in a gyrotropic PhC and demonstrated the potential use for  magnetically-tunable photonic switch devices. This approach provides a realistic and physically accessible platform for generating and manipulating the higher order topology of the AXI PhC, enabling effective topological switching between different axionic hinge of light configurations. In addition to its fundamental theoretical significance, related to the possibility of coupling between photonic axionic excitations and dark-matter axions, the realization of AXI PhC has the potential to open up the field of axion-based topology,  enabling more efficient and versatile control of light propagation in photonic crystals, and thus advancing the state-of-the-art in photonic communication and optical technologies.

\section{Methods}

\textbf{Transversality-Enforced Tight-Binding model for a supercell-modulated PhC.}
To efficiently simulate the electromagnetic response of the photonic \textit{r}AXI, we develop a transversality-enforced tight-binding model (TETB) \cite{antonio2023}, capable of capturing and regularizing the $\Gamma$-point electromagnetic obstruction that arises due to the transversality constraint of the Maxwell equations \cite{antonio2023,christensen2022location}. The model allows the demonstration of HOTI bulk-hinge correspondence of the \textit{r}AXI through cost-effective calculations of large-scale slab- and rod-geometries. 
The underlying PhCs that constitute the starting point of our \textit{r}AXI design are the gyrotropic Weyl semimetals shown of Fig. \ref{weyl}\textcolor{red}{(a)}. Before the introduction of the $z$-directed external magnetic field and SM, the crystal structure belongs to space group (SG) $\#224$ ($Pn\bar{3}m$)  \cite{devescovi2021cubic,devescovi2022vectorial,antonio2023}.

The symmetry content of the photonic bands can be deduced by analyzing the Bloch electric modes (\textbf{E}), obtained in MPB via numerical solution of the Maxwell equations. The \textbf{E} field transforms as a vector:
\begin{equation}
\renewcommand{\theequation}{M.\arabic{equation}} 
\setcounter{equation}{0}
g \textbf{E}(\textbf{r}) = (R\textbf{E})( R^{-1}(\textbf{r}-\textbf{t})),
\end{equation}
for each space group operation $g=\{R|\textbf{t}\}$, where $R$ is a point group element and $\textbf{t}$ a translation.

For each $n$ band with $\omega\neq 0$ and every high-symmetry point $\mathbf{k}_h$, we compute the $x_{n,\mathbf{k}_h}(g)$ diagonal elements of the representation matrix corresponding to $g$ in the little group of $\mathbf{k}_h$, from the overlap integrals:
\begin{equation}
\renewcommand{\theequation}{M.\arabic{equation}} 
x_{n,\mathbf{k}}(g)=\braket{\mathbf{E}_{n,\mathbf{k}_h}|g\mathbf{D}_{n,\mathbf{k}_h}}
\label{transverse_si}
\end{equation}
where $\mathbf{D}=\varepsilon\mathbf{E}$ is the displacement field and $\varepsilon$ the dielectric constant.  From the Schur's Orthogonality Relations \cite{miller1973symmetry}, we are able to extract the symmetry vector $\mathbf{v}^T$ that gives the multiplicity irreducible representation (irrep) in the little group of each high symmetry point. We label the irrep accordingly to the notation of Bilbao Crystallographic Server (BCS) \cite{elcoro2017double}

This analysis returns, for the six lowest-electromagnetic modes: 
\begin{equation}
\renewcommand{\theequation}{M.\arabic{equation}} 
 \begin{aligned}
 \mathbf{v}^T= [&(\blacksquare)^{2T}+\Gamma_2^-+\Gamma_4^-,R_4^-+R_5^+,\\
 &M_1+2M_4,X_1+X_3+X_4]
 \end{aligned}
\label{symVec224}
\end{equation} 
where $(\blacksquare)^{2T}$ indicates the irregular symmetry content at $\Gamma$ and $\omega=0$ arising from transversality of the electromagnetic waves \cite{christensen2022location,antonio2023}, with $(\nonumber)^{T}$ labeling the transverse bands.

A symmetry-constrained, tight-binding Hamiltonian $H(\textbf{k})$ can be constructed for these transverse photonic bands, via the TETB methods proposed by Ref. \cite{antonio2023}. This approach proceeds with the introduction of auxiliary longitudinal modes $\mathbf{v}^L$, able to regularize the $\Gamma$-point obstruction, in order for $\mathbf{v}^{T+L}=\mathbf{v}^T+\mathbf{v}^L$ to be regular. By exploiting a formal mapping between the Schrödinger and electromagnetic wave equations, which relates energies and frequencies quadratically ($\lambda\sim\omega^2$, see \cite{de2018}),
a TETB model is developed enforcing the lowest set of longitudinal bands at $\omega^2 \leq 0$, resulting in the $\mathbf{v}^T=\mathbf{v}^{T+L}-\mathbf{v}^L$ transverse vector capturing all the symmetry, topology and energetic features of the active bands in the PhC.
For the specific $\mathbf{v}^T$ of Eq. \textcolor{red}{M}\ref{symVec224}, this can be achieved via
\begin{equation}\label{eq:GEBR}
\renewcommand{\theequation}{M.\arabic{equation}} 
    \mathbf{v}^{T+L}=A_{2u}@4b+A_{2u}@4c.
\end{equation}
with $\mathbf{v}^L=A_{1}@2a$,  where the decomposition is done in terms of elementary band representations (EBRs), which constitute the trivial atomic limits induced by a localized orbital at a specific Wyckoff position, as in the notation of BCS. This results in a $8$-band model, from $A_{2u}$ photonic pseudo-orbitals at Wyckoff position $4b$ and $4c$.
Gyrotropy can be as well modeled via non-minimal coupling to an external magnetic field $\textbf{H}$:
\begin{equation}
\renewcommand{\theequation}{M.\arabic{equation}} 
    H (\textbf{k},\textbf{H}) = H(\textbf{k}) + f(\textbf{k},\textbf{H}),
\end{equation}
where the function $f(\textbf{k},\textbf{H})$ should respect the symmetries of the crystal, $\textbf{H}$ transforming as a pseudovector. Non-minimal coupling is adopted, due to the uncharged nature of photons, which prevents the use of Peierls substitution. 
The $\textbf{H}=(0,0,H_z)$ field is tuned in order for a Weyl dipole to form along the $k_z$ line, with a separation of $|\mathbf{Q}|=|\mathbf{q}_{+}-\mathbf{q}_{-}|=2\pi/N$ and $N\in\mathds{N}$ and $N\geq2$, as shown in Fig. \ref{weyl}\textcolor{red}{(a)}.

Starting from the $H(\textbf{k},\textbf{H})$ magnetic Hamiltonian, we consider an additional perturbation aiming to capture the effect of a SM of the dielectric elements. The perturbation is introduced as a $z$-periodic on-site potential of Wyckoff positions $4b$ and $4c$:

\begin{equation}\label{eq:mod}
    H_{\Delta}(\textbf{r},\textbf{H}) = H(\textbf{r},\textbf{H}) + \sum_i V_i\cos\left(\frac{2\pi z_i}{N|a|}+\phi\right)c_i^\dagger(\textbf{r}) c_i(\textbf{r}),
\end{equation}
where $H(\textbf{r},\textbf{H})$ is the real-space TETB Hamiltonian for the magnetic system before modulation, $|a|$ is the lattice parameter of the crystal before SM, and $V_i$ and $\phi$ parameterize the amplitude and the phase of the modulation, respectively. Note that the sum in Eq. \eqref{eq:mod} runs over all the basis pseudo-orbitals used in the TETB model, i.e., the pseudo-orbitals placed at WPs $4c$ and $4b$. Since this positions are related by symmetry, the amplitude of the modulation in the positions inside a WP should be equal. We will call them $V_{4c}$ and $V_{4b}$, respectively.

However, since maximal Wyckoff position $4b$ and $4c$ cannot be adiabatically deformed into each other without breaking the symmetry of the model, we have the additional freedom of choosing the relative sign of their modulation amplitude, $V_{4b}$ and $V_{4c}$. Justified by the fact that Wyckoff positions $4c$ fall inside the dielectric elements, while Wyckoff positions $4b$ are in the air region, we decide to adopt the convention where the on-site potentials on $4b$ and $4c$ are opposite in sign, i.e. $V_{4c}=-V_{4b}>0$, consistent with regions of higher and lower electromagnetic energy concentration. As shown in Fig. \ref{weyl}\textcolor{red}{(c)}, the effect of the SM is correctly captured by the transverse modes of the TETB after the introduction of the on-site potential, which results in the opening of a $C_z=1$ gap. 

\textbf{TETB symmetry vectors, double-band inversion and symmetry-constrained $\Gamma$-content}
As we will demonstrate now, the supercell-modulated pseudo-orbitals of the TETB induce all the irreps of the supercell-modulated PhC band-structure, representing an exact representation for the $\Tilde{\mathbf{v}}^{T}_{\phi}$ electromagnetic modes bellow the gap, $\Tilde\nonumber$ standing for the symmetry vector after modulation. Note that we express the symmetry vector in the notation of  MSG $\#2.4$ 
($P\bar{1}$),
which is the symmetry of the crystal after the introduction of the $H_z$ magnetic bias, the $\mathcal{I}$-symmetric SM and the off-axis $h_{x,y}$ perturbation. For the geometry-modulated PhCs, we find:

\begin{equation}
\renewcommand{\theequation}{M.\arabic{equation}} 
 \begin{aligned}
\Tilde{\mathbf{v}}^T_{\phi=0}
= [(\blacksquare)^{2T} +2\Gamma_1^{+}+2\Gamma_1^{-},2R_1^{+}+4R_1^{-},\\3T_1^{+}+3T_1^{-}, 3U_1^{+}+3U_1^{-},2V_1^{+}+4V_1^{-},\\3X_1^{+}+3X_1^{-},3Y_1^{+}+3Y_1^{-},3Z_1^{+}+3Z_1^{-}] 
 \end{aligned}
\end{equation} 
and
\begin{equation}
\renewcommand{\theequation}{M.\arabic{equation}} 
 \begin{aligned}
\Tilde{\mathbf{v}}^T_{\phi=\pi}
= [(\blacksquare)^{2T} +2\Gamma_1^{+}+2\Gamma_1^{-},2R_1^{+}+4R_1^{-},\\3T_1^{+}+3T_1^{-}, 3U_1^{+}+3U_1^{-},4V_1^{+}+2V_1^{-},\\3X_1^{+}+3X_1^{-},3Y_1^{+}+3Y_1^{-},3Z_1^{+}+3Z_1^{-}]
 \end{aligned}
\end{equation}

On the other side, for the onsite-modulated TETB, we obtain:

\begin{equation}
\renewcommand{\theequation}{M.\arabic{equation}} 
 \begin{aligned}
\Tilde{\mathbf{v}}^{T+L}_{\phi=0}
= [4\Gamma_1^{+}+8\Gamma_1^{-},5R_1^{+}+7R_1^{-},\\6T_1^{+}+6T_1^{-}, 6U_1^{+}+6U_1^{-},5V_1^{+}+7V_1^{-},\\6X_1^{+}+6X_1^{-},6Y_1^{+}+6Y_1^{-},6Z_1^{+}+6Z_1^{-}]
 \end{aligned}
\end{equation} 
and
\begin{equation}
\renewcommand{\theequation}{M.\arabic{equation}} 
 \begin{aligned}
\Tilde{\mathbf{v}}^{T+L}_{\phi=\pi}
= [4\Gamma_1^{+}+8\Gamma_1^{-},5R_1^{+}+7R_1^{-},\\6T_1^{+}+6T_1^{-}, 6U_1^{+}+6U_1^{-},7V_1^{+}+5V_1^{-},\\6X_1^{+}+6X_1^{-},6Y_1^{+}+6Y_1^{-},6Z_1^{+}+6Z_1^{-}]
 \end{aligned}
\end{equation}

The TETB therefore correctly models the double band inversion occurring at the $V=(\pi,\pi,0)$ point, between the system with $\phi=0$ and $\phi=\pi$.

After having identified the irregular irrep content at $\Gamma$, as $(\blacksquare)^{2T}=-\Gamma_1^++3\Gamma_1^-$, consistent with symmetry-constrained decomposition for point group $\bar{1}$ as in Refs. \cite{christensen2022location,antonio2023}, we can split the TETB as follows:
$\Tilde{\mathbf{v}}^{T+L}_{\phi}=\Tilde{\mathbf{v}}^T_{\phi}+\Tilde{\mathbf{v}}^L_{\phi}$,
where:
\begin{equation}
\renewcommand{\theequation}{M.\arabic{equation}} 
 \begin{aligned}
\Tilde{\mathbf{v}}^L_{\phi=0,\pi}=[3\Gamma_1^{+}+3\Gamma_1^{-},3R_1^{+}+3R_1^{-},\\3T_1^{+}+3T_1^{-}, 3U_1^{+}+3U_1^{-},3V_1^{+}+3V_1^{-},\\3X_1^{+}+3X_1^{-},3Y_1^{+}+3Y_1^{-},3Z_1^{+}+3Z_1^{-}]
\label{longitudinal_bandrep}
 \end{aligned}
\end{equation}

represents the longitudinal auxiliary modes with ${\omega^2<0}$, and has the same expression for both $\phi=0,\pi$. This shows that the symmetry vector of the TETB represents a precise representation of the electromagnetic modes below the gap of the \textit{r}AXI. Specifically, the TETB symmetry vector with an onsite supercell-modulation can be decomposed as a longitudinal component $\Tilde{\mathbf{v}}^L_\phi$ which has same expression for both $\phi=0,\pi$ phases and a transverse part $\Tilde{\mathbf{v}}^T_\phi$, which coincides with symmetry vector of the transverse modes of the PhC. As we will verify in the next section, $\Tilde{\mathbf{v}}^L_\phi$ has trivial SI, so that the SI of the TETB model and of the MPB solutions coincide.

\textbf{Magnetic symmetry-indicators for the $\mathcal{I}$-invariant photonic \textit{r}AXI.}
In order to assess the role of $\mathcal{I}$-symmetry in the quantization of an \textit{relative}  axion angle, we apply the methods of topological quantum chemistry (TQC) for non-fermionic systems \cite{antonio2023,christensen2022location,manes2020fragile,gutierrez2023topological,xu2022catalogue,de2019engineering} and analyze the symmetry indicators of the modulated PhCs, as the SM angle $\phi$ is varied. In particular, we consider the structures with $\phi=0$ and $\phi=\pi$, in presence of both $H_z$ and a small in-plane $h_{x,y}$ which reduce the symmetry to MSG $2.4$ (in the BNS notation of Refs. \cite{bradley2010mathematical,belov1957neronova}), and we evaluate the  $\{\bar{z}_{2,x},\bar{z}_{2,y},\bar{z}_{2,z}|\bar{z}_4\}$ magnetic SI \cite{watanabe2018structure,kim2019glide,takahashi2020bulk,xu2020high,elcoro2021magnetic}. 

For the effective photonic TETB, which is regular and does not present any obstruction at $\Gamma$, the calculation of the SI follows directly from the well-known closed-formula expression that relates the $\mathcal{I}$-eigenvalues to the $\{\bar{z}_{2,x},\bar{z}_{2,y},\bar{z}_{2,z}|\bar{z}_4\}$ magnetic SI \cite{watanabe2018structure,kim2019glide,takahashi2020bulk,xu2020high,elcoro2021magnetic}, i.e.:
\begin{equation}
\renewcommand{\theequation}{M.\arabic{equation}} 
    \bar{z}_{2,i}= \frac{1}{2}\sum_{\scriptsize\begin{aligned} \mathbf{k}_h \in \{IIMS\} \\[-4pt] \mathbf{k}_{h}\cdot \mathbf{R}_i=\pi\end{aligned}} (n^{+}_h-n^{-}_h) \hspace{0.3cm} \text{mod} \hspace{0.1cm} 2
\end{equation}
\begin{equation}
\renewcommand{\theequation}{M.\arabic{equation}} 
    \bar{z}_{4}= \frac{1}{2}\sum_{\scriptsize \mathbf{k}_h \in \{IIMS\}} (n^{+}_h-n^{-}_h) \hspace{0.3cm} \text{mod} \hspace{0.1cm} 4
\end{equation}
where $n^{+}_h$ ($n^{-}_h$) are the multiplicities of the positive (negative) parity eigenvalues at the high-symmetry point $\mathbf{k}_h$, and $\mathbf{R}_i$ are the primitive lattice vectors. This returns, depending on the phase $\phi$:
\begin{equation}
\renewcommand{\theequation}{M.\arabic{equation}} 
    \nu^{T+L}_{\phi=0}=\{0,0,1|0\}
\end{equation}
and
\begin{equation}
\renewcommand{\theequation}{M.\arabic{equation}} 
    \nu^{T+L}_{\phi=\pi}=\{0,0,1|2\}.
\end{equation}

To obtain the corresponding $ \nu^{T}_{\phi}$ transverse SI for the electromagnetic modes, we can exploit the linearity of the SI with respect to the symmetry vector \cite{cano2021band,christensen2022location}, i.e.:
\begin{equation}
\renewcommand{\theequation}{M.\arabic{equation}} 
 \nu_i^T=\nu_i^{L+T}-\nu_i^{L}.
\end{equation}
Since the SI of the longitudinal modes of Eq. \textcolor{red}{M}\ref{longitudinal_bandrep} are trivial, it follows that the SI of the TETB and the MPB calculations coincide, $\nu_i^T=\nu_i^{L+T}$.
This confirms that the $\phi=0$ and the $\phi=\pi$ systems are obstructed with respect to each other, with a $\delta \bar{z}_4=2$ discontinuity of the \textit{even} $\bar{z}_4$ signaling \textit{relative} axion topology.

\textbf{Acknowledgements} \par 
A.G.E., A.M.P and C.D. acknowledges support from the Spanish Ministerio de Ciencia e Innovación (PID2019-109905GA-C2) and by the Gipuzkoa Provincial 
Council within the QUAN-000021-01 project. M.G.D., R.I., A.M.P. and M.G.V. acknowledge   the   Spanish   Ministerio  de  Ciencia  e  Innovacion  (grant PID2019-109905GB-C21). 
A.G.E. and M.G.V. acknowledge funding from the IKUR Strategy under the collaboration agreement between Ikerbasque Foundation and DIPC on behalf of the Department of Education of the Basque Government, Programa de ayudas de apoyo a los agentes de la Red Vasca de Ciencia, Tecnolog\'ia e Innovaci\'on acreditados en la categor\'ia de Centros de Investigaci\'on B\'asica y de Excelencia (Programa BERC) from the Departamento de Universidades e Investigaci\'on del Gobierno Vasco and Centros Severo Ochoa AEI/CEX2018-000867-S from the Spanish Ministerio de Ciencia e Innovaci\'on.
M.G.V. thanks support to the Deutsche Forschungsgemeinschaft (DFG, German Research Foundation) GA 3314/1-1 – FOR 5249 (QUAST) and partial support from European Research Council (ERC) grant agreement no. 101020833.  The work of JLM has been  supported in part by the Basque Government Grant No. IT1628-22 and the PID2021-123703NB-C21 grant funded by MCIN/AEI/10.13039/501100011033/ and by ERDF; ``A way of making Europe”. The work of B.B. and Y.~H. is supported by the Air Force Office of Scientific Research under award number FA9550-21-1-0131. Y.~H. received additional support from the US Office of Naval Research (ONR) Multidisciplinary University Research
Initiative (MURI) grant N00014-20-1-2325 on Robust
Photonic Materials with High-Order Topological Protection. C.D. acknowledges financial support from the MICIU through the FPI PhD Fellowship CEX2018-000867-S-19-1. M.G.D. acknowledges financial support from Government of the Basque Country through the pre-doctoral fellowship PRE\_2022\_2\_0044.
\bibliographystyle{unsrt}
\bibliography{biblio}

\end{document}